\documentclass[sigconf]{acmart}

\usepackage{xspace}
\usepackage{hyperref}
\usepackage{bm} 
\usepackage{subfig}
\usepackage{multirow}
\newcommand{\llmts}{LLMTS\xspace} 

\newcommand{\modelname}{Waltz\xspace}
\newcommand{\modelnamenospace}{Waltz}

\AtBeginDocument{%
  }

\setcopyright{acmlicensed}
\copyrightyear{2018}
\acmYear{2018}
\acmDOI{XXXXXXX.XXXXXXX}

\acmConference[Conference acronym 'XX]{Make sure to enter the correct
  conference title from your rights confirmation email}{June 03--05,
  2018}{Woodstock, NY}
\acmISBN{978-1-4503-XXXX-X/2018/06}

\begin{document}

\title{Watermarking Large Language Model-based Time Series Forecasting}

\author{Wei Yuan}
\email{w.yuan@uq.edu.au}
\affiliation{%
  \institution{The University of Queensland}
  \city{Brisbane}
  \state{QLD}
  \country{Australia}
}

\author{Chaoqun Yang}
\email{chaoqun.yang@griffith.edu.au}
\affiliation{%
  \institution{Griffith University}
  \city{Gold Coast}
  \state{QLD}
  \country{Australia}
}

\author{Yu Xing}
\email{wake.xingyu@gmail.com}
\affiliation{%
  \institution{Nanjing University}
  \city{Nanjing}
  \state{Jiangsu}
  \country{China}
}

 \author{Tong Chen}
\email{tong.chen@uq.edu.au}
\affiliation{%
  \institution{The University of Queensland}
  \city{Brisbane}
  \state{QLD}
  \country{Australia}
}

\author{Nguyen Quoc Viet Hung}
\email{henry.nguyen@griffith.edu.au}
\affiliation{%
  \institution{Griffith University}
  \city{Gold Coast}
  \state{QLD}
  \country{Australia}
}

\author{Hongzhi Yin}\authornote{Corresponding author.}
\email{h.yin1@uq.edu.au}
\affiliation{%
  \institution{The University of Queensland}
  \city{Brisbane}
  \state{QLD}
  \country{Australia}
}
\renewcommand{\shortauthors}{Yuan et al.}

\begin{abstract}
Large Language Model-based Time Series Forecasting (\llmts) has shown remarkable promise in handling complex and diverse temporal data, representing a significant step toward foundation models for time series analysis. However, this emerging paradigm introduces two critical challenges. 
First, the substantial commercial potential and resource-intensive development raise urgent concerns about intellectual property (IP) protection. 
Second, their powerful time series forecasting capabilities may be misused to produce misleading or fabricated deepfake time series data.
To address these concerns, we explore watermarking the outputs of \llmts models, that is, embedding imperceptible signals into the generated time series data that remain detectable by specialized algorithms.
We propose a novel post-hoc watermarking framework, \modelname, which is broadly compatible with existing \llmts models.
\modelname is inspired by the empirical observation that time series patch embeddings are rarely aligned with a specific set of LLM tokens, which we term ``cold tokens''. Leveraging this insight, \modelname embeds watermarks by rewiring the similarity statistics between patch embeddings and cold token embeddings, and detects watermarks using similarity z-scores.
To minimize potential side effects, we introduce a similarity-based embedding position identification strategy and employ projected gradient descent to constrain the watermark noise within a defined boundary.
Extensive experiments using two popular \llmts models across seven benchmark datasets demonstrate that \modelname achieves high watermark detection accuracy with minimal impact on the quality of the generated time series. 
\end{abstract}

\ccsdesc[500]{Mathematics of computing~Time series analysis}
\ccsdesc[300]{Social and professional topics~Copyrights}

\keywords{Time Series Forecasting, Large Language Model, Digital Watermarking}


\maketitle

\section{Introduction}\label{sec_intro}
Time series forecasting plays a crucial role in a wide range of real-world applications, including finance, healthcare, weather prediction, and energy management~\cite{lim2021time}. 
Over the years, substantial efforts have been dedicated to improving forecasting accuracy through incorporating various advanced neural architectures, such as CNNs~\cite{liu2022scinet}, GNNs~\cite{jin2024survey}, MLPs~\cite{zeng2023transformers}, and Transformers~\cite{wen2023transformers}.

Recently, with growing recognition of the formidable capabilities of Large Language Models (LLMs) in processing sequential data~\cite{min2023recent}, an increasing number of studies have begun to repurpose LLMs for time series forecasting~\cite{jin2024position,zhang2024large}, giving rise to a new paradigm we refer to as LLM-based Time Series Forecasting (\llmts).
In a typical \llmts pipeline~\cite{jintime,zhou2023one,cao2024tempo,liu2024unitime}, raw time series data is first projected, either explicitly or implicitly, into the LLM's embedding space using a suitable encoder.
The encoded sequence, often wrapped in carefully designed prompts, is then processed by the LLM, and the output is subsequently decoded to generate the forecast.
For example, Zhou et al.~\cite{zhou2023one} demonstrated that by linearly projecting time series patches and fine-tuning only the positional embeddings and normalization layers, a pretrained LLM can achieve state-of-the-art performance while exhibiting strong few-shot and zero-shot forecasting capabilities.
Building on this insight, TEMPO~\cite{cao2024tempo} introduced signal decomposition and prompt engineering to further enhance the model's temporal understanding. 
UniTime~\cite{liu2024unitime} adopted a masking-based training objective to develop a unified forecasting model across multiple datasets. 
Other recent works~\cite{jintime} have explored more refined alignment strategies between time series and natural language representations to improve foundational forecasting performance.

\begin{figure}[!t]
  \centering
\includegraphics[width=0.8\linewidth]{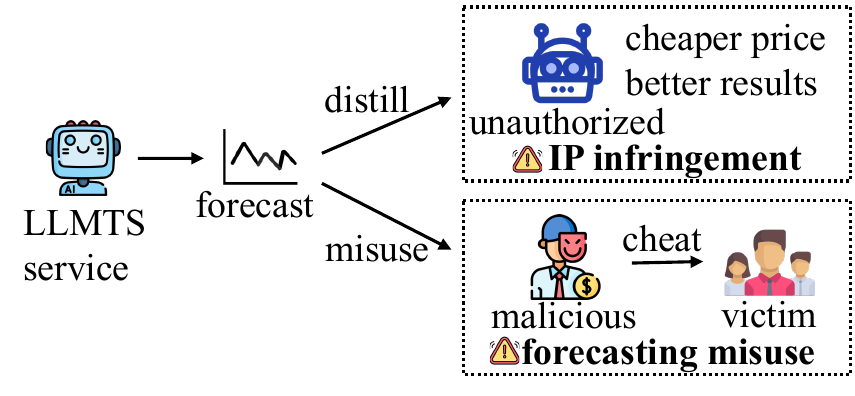}
  \caption{The threats for commercial \llmts service.}\label{fig_threats}
\end{figure}

The growing success of \llmts models and their foundational capabilities has already demonstrated substantial commercial potential. However, we argue that two critical risks have been ignored, as shown in Figure~\ref{fig_threats}.
Firstly, the development of \llmts involves considerable investment in model design and training, often largely exceeding the cost of traditional forecasting models, which raises concerns around model intellectual property (IP) protection.
A common practice is to deploy these models as black-box services, offering access solely through APIs (i.e., the model-as-a-service paradigm). 
While this commercial model appears secure, malicious users can still exploit it by querying the service and distilling a similar model at a small fraction of the original cost, thereby undermining the model's market value.
Second, the powerful generative capabilities of \llmts models introduce potential risks of misuse. For instance, dishonest users could employ these models to generate fake yet convincing time series data, such as manipulated stock price trends or synthetic trading volumes, with the intent to deceive investors or mislead automated systems.

In computer vision and natural language processing, digital watermarking has become a widely adopted method for protecting model intellectual property and tracing the misuse of generated content~\cite{kirchenbauer2023watermark,zhang2020model,wan2022comprehensive,liu2024survey}. 
These approaches embed imperceptible yet statistically verifiable signals into model outputs, allowing owners to detect unauthorized use without compromising utility. 
This success naturally motivates the exploration of watermarking for large language model-based time series forecasting (\llmts) to address similar risks. 
However, directly transferring existing watermarking techniques from vision or text domains to \llmts is far from straightforward due to fundamental differences in data characteristics, model structures, and task requirements. 
Specifically, time series data consist of continuous-valued sequences, making methods designed for multi-channel images or discrete tokens in natural language unsuitable for direct application. 
Furthermore, many established approaches embed watermarks during training by modifying loss functions, architectures, or decoding strategies, whose steps that are impractical for \llmts, where training is computationally expensive and often inaccessible to service providers maintaining already-deployed models. 
Beyond these domain-specific obstacles, watermarking \llmts must also satisfy general requirements common to ``good" watermarks in other areas: watermarks must remain imperceptible while being reliably detectable and resist under threats such as unauthorized model distillation.
All these challenges make watermarking \llmts a non-trivial task.

In this paper, we propose \modelname, a novel, generic, and robust watermarking framework for large language model-based time series forecasting. 
\modelname is designed as a post-hoc framework that requires only the model's output and limited prior knowledge of the protected \llmts model, making it broadly generalizable to most \llmts implementations.
To ensure that the watermark is both detectable and minimally intrusive to forecasting performance, \modelname embeds watermarks by enhancing the semantic alignment between output time series patch embeddings and a set of ``cold'' natural language tokens.
Specifically, at the initial stage, \modelname constructs a cold-token-based watermark book by analyzing the alignment statistics.
Then, for a time series data, it first identifies the most suitable watermark embedding locations based on similarity metrics to minimize the side effects of semantic rewiring.
After that, a projected gradient descent optimization method is employed to optimize the watermarking objective while constrain the scale of watermark noise, providing a guaranteed upper bound on performance degradation due to watermarking. 
Finally, a training-free, z-score-based detection module is introduced to effectively identify watermarked time series data.
We validate \modelname on two representative \llmts models and seven time series forecasting benchmarks across four domains. Our results demonstrate that \modelname achieves high watermark detection accuracy with minimal impact on forecasting performance.
Moreover, to demonstrate the robustness of the watermark, we distill models based on the watermarked outputs and find that the output of these distilled models continue to exhibit the watermark signal, confirming the traceability of the method under realistic threat models.

In summary, this paper makes the following contributions:
\begin{itemize}
  \item We identify two key risks, intellectual property leakage and misuse of generated content, that threaten the commercial viability of \llmts and motivate the need for watermarking.
  \item We propose \modelname, the first generic, post-hoc watermarking framework for LLM-based time series forecasting, which ensures watermark robustness and utility preservation.
  \item Extensive experiments on seven time series forecasting datasets covering four different domains with two typical \llmts models showcase the superiority of our \modelname.
\end{itemize}

\section{Related Work}\label{sec_relatedwork}
In this section, we briefly review the literature on two closely related topics: LLM for time series forecasting and digital watermarking.
More related works can be found in Appendix~\ref{apd_relatedwork}.

\subsection{LLM for Time Series Forecasting}
LLMs have been applied to time series forecasting in two main ways: leveraging text information for multimodal forecasting, and directly modeling time series data using LLMs. This paper focuses on the latter, given its rapid progress. Early work by Gruver et al.~\cite{gruver2023large} demonstrated LLMs' zero-shot capabilities on time series. Zhou et al.~\cite{zhou2023one} later provided theoretical support, showing that self-attention resembles principal component analysis. Since then, many \llmts methods have emerged. A representative framework is TEMPO~\cite{cao2024tempo}, which uses time series-specific encoders and decoders around an LLM. This architecture is widely adopted by subsequent works such as UniTime~\cite{liu2024unitime}, TimeLLM~\cite{jintime}, and others~\cite{pan2024s,jin2024position}

\subsection{Digital Watermarking}
Digital watermarking~\cite{cox2002digital} is widely used to protect model copyrights and prevent the misuse of synthetic content in generative AI~\cite{wen2023tree,kirchenbauer2023watermark,liang2024watermarking,zhang2024remark}. As \llmts advance, they face similar risks of unauthorized use and IP infringement as models in vision and NLP. However, watermarking research in \llmts remains scarce. To date, HTW~\cite{Schaik2025Robust} is the only method specifically designed for this setting, but it is simplistic and significantly degrades forecasting accuracy, as shown in Table~\ref{tb_watermark_validate}. This motivates the need for a more robust and effective watermarking framework.

\section{Preliminaries}\label{sec_preliminaries}
\subsection{LLM-based Time Series Forecasting}
Given the observed previous $K$ timestamps $x_{t-K}^{i}, \dots, x_{t-1}^{i}$, the task of LLM-based time series forecasting aims to predict the values for the next $L$ timestamps, formally:
\begin{equation}\label{eq_llmts}
  \hat{x}_{t}^{i},\dots,\hat{x}_{t+L-1}^{i} = \mathcal{F}(x_{t-K}^{i}, \dots, x_{t-1}^{i}; \mathcal{P}; \mathbf{\Theta})
\end{equation}
where $\hat{x}_{t}^{i},\dots,\hat{x}_{t+L-1}^{i}$ are the predicted $L$-timestamps future values for channel $i$, $F$ is the \llmts model with parameters $\mathbf{\Theta}$, and $\mathcal{P}$ is the prompt used to describe the task, specific domain or the characteristics of the data. 
Note that most existing \llmts works~\cite{zhou2023one,cao2024tempo,liu2024unitime,jintime,pan2024s} treat the multivariate time series as multiple independent univariate time series according to channel independence~\cite{nietime}.
Therefore, in the following presentation, we will omit the channel index $i$ to keep concise.

Generally, the detailed process of E.q.~\ref{eq_llmts} is as follows. The \llmts model $F$ will fist segment the input time series $x_{t-K}, \dots, x_{t-1}$ into a sequence of patches $[\mathbf{x}_{j}]_{j=1}^{N}$ that $N=\lceil \frac{K-P}{S}\rceil + 1$ is the number of patches, $\mathbf{x}_{j}\in\mathbb{R}^{1\times P}$ is the $j$-th patch, and $P$ is the patch length while $S$ is the stride length.
Then, these patches are feed into certain time series encoder $\mathcal{E}^{ts}$ to obtain a sequence of time series embeddings $[\mathbf{e}_{j}^{ts}]_{j=1}^{N}$.
These embeddings are wrapped with prompt $\mathcal{P}$'s token embeddings to feed to a LLM model to get a sequence of latent vectors $[\mathbf{h}_{j}^{ts}]_{j=1}^{N}$.
Based on these latent vectors, \llmts adopts certain time series decoder $\mathcal{D}^{ts}$ to get the estimated future values $\hat{x}_{t},\dots,\hat{x}_{t+L-1}$.

In this paper, to demonstrate the generalization of our watermarking framework, we use two well-known and representative open-sourced \llmts (TEMPO~\cite{cao2024tempo} and UniTime~\cite{liu2024unitime}) as the base \llmts models.

\begin{figure*}[h]
  \centering
  \includegraphics[width=0.86\linewidth]{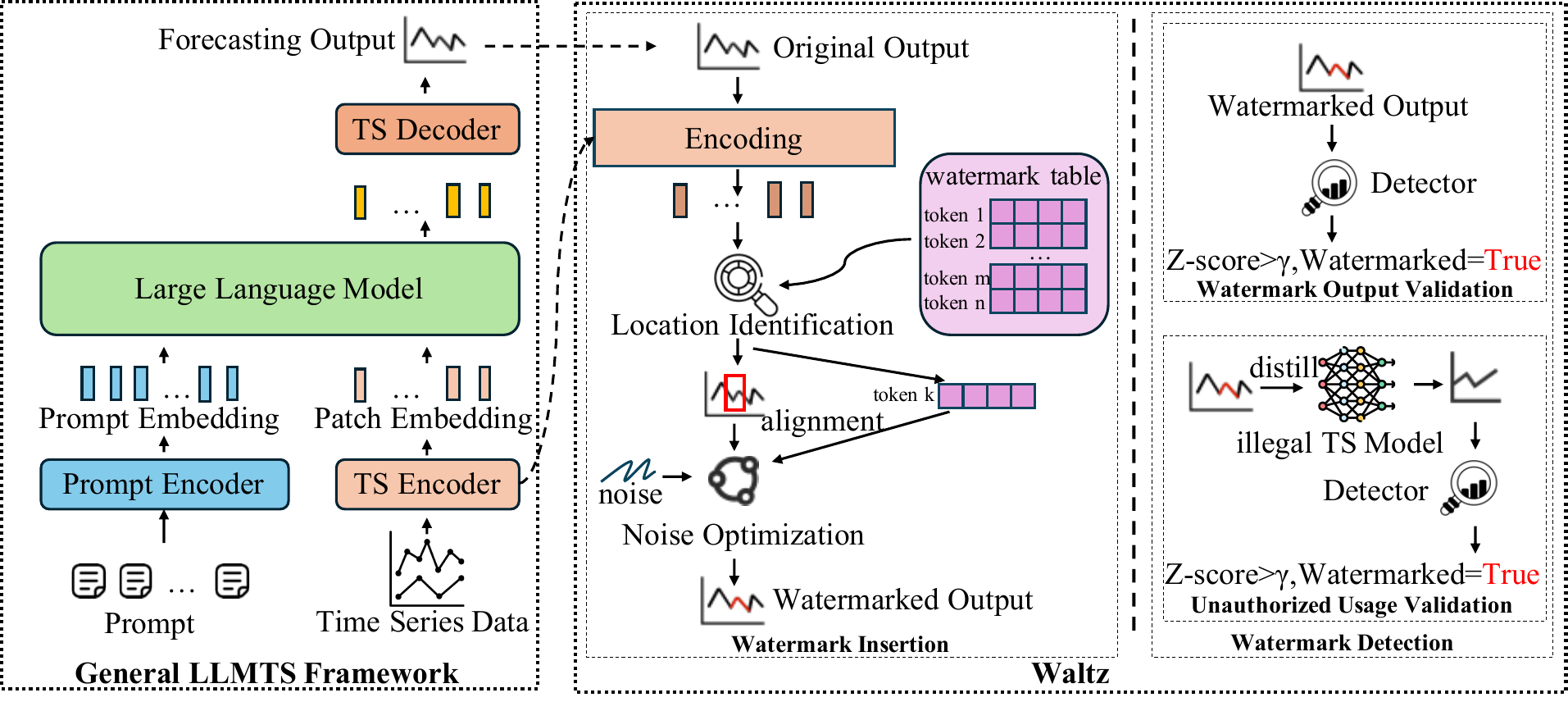}
  \caption{The general \llmts framework and the overview of our \llmts watermark framework.}\label{fig_overview}
\end{figure*}

\begin{figure}[!h]
  \centering
  \subfloat[TEMPO.]{\includegraphics[width=0.23\textwidth]{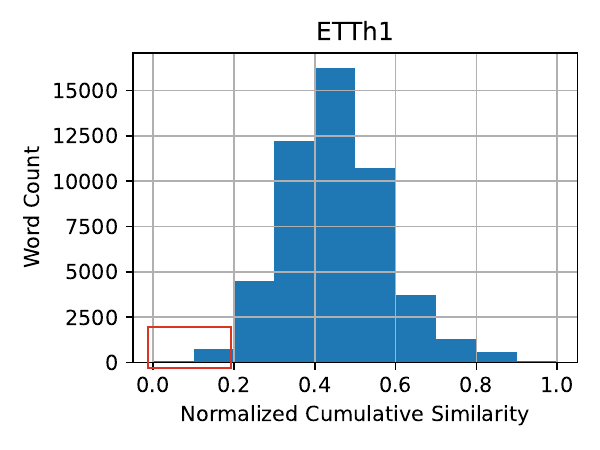}\label{fig_align_tempo}}
  \hfil
  \subfloat[UniTime]{\includegraphics[width=0.23\textwidth]{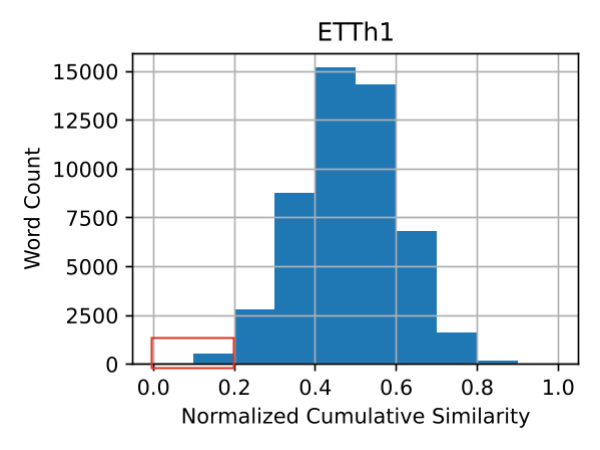}\label{fig_align_unitime}}
  \caption{Proof-of-Concepts: The distribution of natural language tokens' accumulated similarity with time patches on ETTh1. There are a set of tokens that are dissimilar with patches, namely ``cold tokens'' (highlighted by red box). Same observation can be obtained from other datasets.}\label{fig_align}
  \vspace{-8pt}
\end{figure}

\subsection{Problem Definition of Watermark for LLMTS}
Assume $\mathcal{F}(\boldsymbol{\Theta})$ is the \llmts model that we aim to protect through watermarking. A watermark embedding algorithm $\mathcal{A}^{e}_{wm}$ is applied to the model's predicted output $\hat{\mathbf{x}}_{t:t+L-1}$, producing a watermarked version $\hat{\mathbf{x}}_{t:t+L-1}^{wm}$. A corresponding watermark detection algorithm $\mathcal{A}^{d}_{wm}$ is then used to identify whether a given time series segment contains a watermark. 
This mechanism can help prevent unauthorized use or distribution of model-generated outputs, protecting the model copyright.
Moreover, a good watermark can even be traceable when using the watermarked data to unauthorizedly train a model.
Specifically, if an adversary misappropriates a set of model outputs $\hat{\mathbf{x}}_{t:t+L-1}$ from $\mathcal{F}(\boldsymbol{\Theta})$ to train a surrogate or distilled model $\mathcal{F}^{distill}$, a well-designed watermark detection algorithm $\mathcal{A}^{d}_{wm}$ should still be able to identify the presence of watermarks in the outputs generated by $\mathcal{F}^{distill}$.

Generally, an effective watermarking scheme should satisfy the following three key properties:
(1) Detectability: The watermark must be reliably detectable by the designated detection algorithm, with low false positives and false negatives.
(2) Traceability: The watermark should remain detectable even after usage, such as output distillation or model extraction. This ensures that unauthorized reuse of the model's predictions can be traced back to the original source.
(3) Imperceptibility: Embedding the watermark should not significantly degrade the performance or utility of the model. The watermarked outputs should remain practically indistinguishable from non-watermarked ones in terms of forecasting quality.
It is relatively easy to satisfy the first two properties by injecting conspicuous or abnormal patterns into the time series data, which would be trivially detectable. However, the primary challenge lies in achieving a balanced trade-off among all three, i.e., embedding watermarks that are both robust and subtle, so they are effective for detection and traceability, yet negligible in impact on model performance.

Beyond effectiveness, the amount of prior knowledge required by a watermarking framework is a critical factor in determining its practical applicability. In this paper, we adopt a strict yet realistic setting in which the watermarking process is executed by a third party with limited access to the protected model. Specifically, the third party can only interact with the model via model-as-a-service APIs, such as the time series encoding service and the model’s output to be watermarked. Under this constrained assumption, our watermarking framework remains broadly applicable to a wide range of \llmts.

\section{Methodology}\label{sec_methodology}
\subsection{Overview}
As illustrated in the left part of Figure~\ref{fig_overview}, Large Language Model-based Time Series models (\llmts) utilize LLMs as their backbone to capture temporal patterns. As a result, the intermediate representations of time series data, i.e., the patch embeddings, naturally align with the LLM's token embedding space during training, either explicitly or implicitly.
This observation inspires our watermarking approach, \modelnamenospace : we can embed watermarks by statistically altering the alignment between time series patch embeddings and LLM token embeddings.
The right part of Figure~\ref{fig_overview} shows the pipeline of \modelname, which consists of two main components: watermark insertion and watermark detection.
In the following subsections, we will describe the technical details of these components.

\subsection{Watermark Insertion}
To embed a watermark by modifying the similarity between patch embeddings and token embeddings, three key questions must be addressed:
\emph{Q1}: Which tokens should the patch embeddings be aligned with?
\emph{Q2}: Which patches should be modified, that is, where should the watermark be embedded?
\emph{Q3}: How can the desired similarity realignment be effectively achieved?
In \modelname, the watermark insertion process is designed to address these questions through three core components: cold token-based watermark book construction for \emph{Q1}, similarity-based watermark location identification for \emph{Q2}, and projected gradient descent-based watermark optimization for \emph{Q3}.

\noindent\textbf{Cold Token-based Watermark Book Construction.}
For \emph{Q1}, a naive approach would be to randomly select a set of tokens to serve as the watermark book and then modify the similarity between patch embeddings and these token embeddings. However, this strategy often leads to suboptimal performance, as it may require large similarity changes to make the watermarked patches statistically distinguishable from normal ones.

Empirical observations suggest a more effective approach. We find that well-trained \llmts models tend to contain a group of ``cold'' tokens, i.e., tokens that consistently exhibit low similarity to all patch embeddings and are rarely aligned with them. This phenomenon is illustrated in Figure~\ref{fig_align} with proof-of-concept results.
Building on this insight, \modelname aims to rewire certain patch embeddings to become abnormally similar to these cold tokens. Because such tokens are typically not aligned with any patches, even a small increase in similarity is statistically significant and can serve as a strong watermark signal. This significantly lowers the difficulty of watermark embedding and minimizes the impact on forecasting performance.
Specifically, \modelname requires the \llmts model that needs to be protected to provide a collection of cold tokens, which are used to construct the watermark book, denoted as $\mathcal{B} = \{w_{i}^{cold}\}_{i=1}^{M}$, and the mean $\mu$ and standard deviation $\sigma$ of the similarity scores which we will use for watermark detection later. The \llmts model identifies these cold tokens by computing the accumulated similarity between patch and token embeddings over a subset of its training data, then selecting the tokens with the lowest accumulated similarity as the cold token set.

\noindent\textbf{Similarity-based Watermark Location Identification.}
This module aims to identify which patch should be watermarked to ensure effective detection (i.e., \emph{Q2}) while minimizing the impact on output quality.
Given a model forecast $\hat{x}_t, \dots, \hat{x}_{t+L-1}$, we first encode it into a set of patch embeddings $\mathbf{E} = \{\hat{\mathbf{e}}_{j}\}_{j=1}^{\lceil \frac{L - P}{S}\rceil}$ by calling the protected \llmts model's encoding service.
We then compute the cosine similarity between each patch embedding and each token in the watermark book $\mathbf{B}$. 
Based on these similarity scores, we select the top $\alpha$ most-aligned patch-token pairs for watermark embedding:
\begin{equation}
  E', \mathbf{B}' \leftarrow argmax (sim(\mathbf{E}, \mathbf{B})), |E'| =\alpha
\end{equation}
Here, $E'$ denotes the selected patch embeddings, and $\mathbf{B}' \in \mathbb{R}^{|E'| \times d}$ represents the corresponding token embeddings from the watermark book that each selected patch will be aligned with (i.e., ``rewired'' toward).
Next, we trace the selected patch embeddings back to their corresponding time series segments, denoted as $\hat{\mathbf{x}}'$, where optimized watermark noise will be added. These segments may be either continuous or non-contiguous. For clarity of presentation, we use $\hat{\mathbf{x}}'$ to collectively refer to all such selected segments of the time series that will receive watermarking.

\noindent\textbf{Projected Gradient Descent-based Watermark Noise Optimization.} 
After determining the rewiring positions and their corresponding target tokens, the final challenge in watermark embedding is how to effectively perform the similarity realignment (i.e., \emph{Q3}).
In \modelname, we perturb the selected time series segments with additive noise $\boldsymbol{\epsilon}$ to make their patch embeddings more similar to the selected cold token embeddings. This process can be formalized as the following optimization objective:
\begin{align}\label{eq_optim_obj_simple}
  L(\boldsymbol{\epsilon}) = (1 - sim(\mathcal{E}^{ts}(\hat{\mathbf{x}}' + \boldsymbol{\epsilon}), \mathbf{B}'))
\end{align}

To minimize the impact on forecasting accuracy, the noise $\boldsymbol{\epsilon}$ must be kept small. A straightforward solution is to apply a regularization term that penalizes the magnitude of the noise:
\begin{equation}\label{eq_optim_with_norm}
  L(\boldsymbol{\epsilon})' = L(\boldsymbol{\epsilon}) + \lambda \left\|\boldsymbol{\epsilon}\right\|
\end{equation}
where $\lambda$ is a hyperparameter that balances alignment strength and noise scale. However, this ``soft constraint'' has two key limitations:
(1) It provides no explicit guarantee on the maximum distortion introduced by $\boldsymbol{\epsilon}$, (2) $\lambda$ typically requires careful tuning for different time series data.

To address these issues, \modelname formulates watermark embedding as a constrained optimization problem:
\begin{align}\label{eq_optim_obj}
  L(\boldsymbol{\epsilon}) = (1 - sim(\mathcal{E}^{ts}(\hat{\mathbf{x}}' + \boldsymbol{\epsilon}), \mathbf{B}')), \left\|\boldsymbol{\epsilon}\right\| \leq \eta
\end{align}
where $\eta$ is a fixed noise budget that strictly bounds the perturbation within $[-\eta, \eta]$, providing a guaranteed upper limit on the watermark's impact.

We solve this constrained optimization problem (E.q.~\ref{eq_optim_obj}) using Projected Gradient Descent (PGD)~\cite{madry2017towards}:
\begin{equation}
  \hat{\boldsymbol{\epsilon}} \leftarrow \mathop{\prod}\limits_{\mathcal{C}_{\infty}(0,\eta)} (\hat{\boldsymbol{\epsilon}} - \beta \nabla L(\boldsymbol{\epsilon}))
\end{equation}
where $\mathop{\prod}\limits_{\mathcal{C}_{\infty}(0,\eta)}$ indicates a projection onto a $l_{\infty}$-norm ball with radius $\eta$ and $\beta$ is the step size.
In practice, the projection operator is implemented via simple element-wise clipping:
\begin{equation}
  \mathop{\prod}\limits_{\mathcal{C}_{\infty}(0,\eta)}(\hat{\boldsymbol{\epsilon}}_{i}) =
\begin{cases}
\hat{\boldsymbol{\epsilon}}, & \text{if } \hat{\boldsymbol{\epsilon}}_{i} \leq \eta \\
sign(\hat{\boldsymbol{\epsilon}}_{i})\eta, & \text{otherwise}
\end{cases}
\end{equation}
It is important to note that, although the noise construction involves optimization process, the computational overhead is minimal due to two factors: 
(1) The number of trainable parameters (i.e.,$\boldsymbol{\epsilon}$ ) equals $P \cdot \alpha$, which is quite small. For example, the default value of $P$ and $\alpha$ are $16$ and $1$ in this paper, resulting in only $16$ float numbers for optimization
(2) Since we only need to statistically shift the patch embedding towards the target token, rather than accurately make them similar, we can set a very large step size $\beta$ with just a few iteration step.

After we obtained an optimized noise $\boldsymbol{\epsilon}$, we add them to the corresponding time series position to form the watermarked data, such as:
\begin{equation}
  \hat{\mathbf{x}}^{wm}_{t:t+L-1} = \hat{x}_{t}, \dots, \hat{x}_{t+j} + \hat{\epsilon}_{0}, \dots, \hat{x}_{t+j+P} + \hat{\epsilon}_{P}, \dots, \hat{x}_{t+L-1}
\end{equation}
where we assume $\hat{x}_{t+j}$ to $\hat{x}_{t+j+P}$ are the segment that we identify to add watermark. 
Note that according to the setting of $\alpha$, the number of segments we watermarked can be various.

\subsection{Watermark Detection}
Given a sample $\widetilde{\mathbf{x}}$, similar to watermark insertion process, we use the \llmts encoder $\mathcal{E}^{ts}$ to obtain patch embeddings.  
Then, we need to identify whether these patch embeddings having been rewired to the tokens in watermark book $\mathcal{B}$.

A naive detection method would be to check whether any cold token from $\mathcal{B}$ appears among the top-$K$ most similar tokens for any patch:
\begin{equation}
  y = topK(sim(\mathcal{E}^{ts}(\widetilde{\mathbf{x}}), \mathbf{V})) \cap \mathcal{B} \neq \emptyset
\end{equation}
where $y$ is the binary detection result indicating whether the time series $\widetilde{\mathbf{x}}$ was generated by the protected \llmts model (i.e., is watermarked), and $\mathbf{V}$ is LLM's token embedding table.
However, this naive approach is highly sensitive to the effectiveness of the similarity realignment. 
Since noise $\boldsymbol{\epsilon}$ is expected to be small, the changes in similarity are often modest, making this method prone to false positives.

To improve robustness, \modelname employs a statistical detection method based on the z-score, which measures how significantly a sample deviates from the normal distribution of similarities between unwatermarked patch embeddings and watermark book tokens.
\begin{equation}
  y = \frac{max(sim(\mathcal{E}^{ts}(\widetilde{\mathbf{x}}), \mathbf{B})) - \mu}{\sigma} > \gamma
\end{equation}
where $\gamma$ is the threshold, $\mu$ and $\sigma$ are statistics that we obtained when the protected \llmts constructs watermark book $\mathcal{B}$. 
Note that we use the maximum similarity because \modelname modifies only a few selected patches within the time series, leaving the majority unchanged. 
Consequently, pooling strategies such as averaging or summing similarity scores will dilute the watermark signal, rendering the statistical change too subtle for reliable detection.

\begin{table*}[!ht]
  \caption{Comparison of watermarked output detection and the impact on utility. Bold value of $\frac{F1}{\Delta MSE}$ indicates achieving better balance in watermark imperceptibility and detectability and Bold Value for MSE and MAE means the best value except for the Clean output. "-" means that we cannot calculate F1 or $\frac{F1}{\Delta MSE}$ scores for Clean.}
\label{tb_watermark_validate}
\begin{tabular}{llccccc|ccccc}
\hline
                                                        &                                      & \multicolumn{5}{c|}{\textbf{TEMPO}}                                       & \multicolumn{5}{c}{\textbf{UniTime}}                                      \\ \hline
   \textbf{Dataset}                                                     &                                      & \textbf{Clean} & \textbf{HTW} & \textbf{FSW} & \textbf{\modelname(soft)} & \textbf{\modelname} & \textbf{Clean} & \textbf{HTW} & \textbf{FSW} & \textbf{\modelname(soft)} & \textbf{\modelname} \\ \hline
\multicolumn{1}{l|}{\multirow{4}{*}{\textbf{ETTh1}}}    & \multicolumn{1}{l|}{\textbf{MSE}$\downarrow$}    & 0.39986      & 0.45723      & 0.44096      & 0.41962      & \textbf{0.40924}       & 0.39479      & 0.45881      & 0.43812      & 0.41061      & \textbf{0.40434}       \\ 
\multicolumn{1}{l|}{}                                   & \multicolumn{1}{l|}{\textbf{MAE}$\downarrow$}    & 0.40471      & 0.45644      & 0.44768      & 0.42415      & \textbf{0.41300}       & 0.41785      & 0.47233      & 0.45939      & 0.43151      & \textbf{0.42661}       \\
\multicolumn{1}{l|}{}                                   & \multicolumn{1}{l|}{\textbf{F1}$\uparrow$}     & -            & 0.78309      & 0.81351      & 0.84910      & \textbf{0.88600}       & -            & 0.79731      & 0.86413      & \textbf{0.94065}      & 0.90891       \\
\multicolumn{1}{l|}{}                                   & \multicolumn{1}{l|}{$\frac{F1}{\Delta MSE}\uparrow$} & -            & 13.650       & 19.793       & 42.971       & \textbf{94.456}       & -            & 12.454       & 19.943       & 59.460       & \textbf{95.174}       \\ \hline
\multicolumn{1}{l|}{\multirow{4}{*}{\textbf{ETTh2}}}    & \multicolumn{1}{l|}{\textbf{MSE}$\downarrow$}    & 0.29457      & 0.34495      & 0.33652      & 0.32126      & \textbf{0.30424}       & 0.29108      & 0.34148      & 0.33596      & 0.30791      & \textbf{0.30009}       \\
\multicolumn{1}{l|}{}                                   & \multicolumn{1}{l|}{\textbf{MAE}$\downarrow$}    & 0.35014      & 0.40718      & 0.41080      & 0.38522      & \textbf{0.36363}       & 0.34849      & 0.40352      & 0.40737      & 0.36858      & \textbf{0.36101}       \\
\multicolumn{1}{l|}{}                                   & \multicolumn{1}{l|}{\textbf{F1}$\uparrow$}     & -            & 0.78636      & 0.65351      & \textbf{0.91847}      & 0.91793       & -            & 0.82450      & 0.65794      & 0.96562      & \textbf{0.97005}       \\
\multicolumn{1}{l|}{}                                   & \multicolumn{1}{l|}{$\frac{F1}{\Delta MSE}\uparrow$} & -            & 15.609       & 15.578       & 34.413       & \textbf{94.926}        & -            & 16.359       & 14.660       & 57.375       & \textbf{107.664}       \\ \hline
\multicolumn{1}{l|}{\multirow{4}{*}{\textbf{ETTm1}}}    & \multicolumn{1}{l|}{\textbf{MSE}$\downarrow$}    & 0.30556      & 0.36243      & 0.34602      & 0.33602      & \textbf{0.31368}       & 0.31908      & 0.38138      & 0.36308      & 0.33240      & \textbf{0.32699}       \\
\multicolumn{1}{l|}{}                                   & \multicolumn{1}{l|}{\textbf{MAE}$\downarrow$}    & 0.34835      & 0.40612      & 0.39918      & 0.38399      & \textbf{0.35895}       & 0.36446      & 0.42302      & 0.41321      & 0.38101      & \textbf{0.37269}       \\
\multicolumn{1}{l|}{}                                   & \multicolumn{1}{l|}{\textbf{F1}$\uparrow$}     & -            & 0.78395      & 0.80536      & 0.86323      & \textbf{0.87137}       & -            & 0.79904      & 0.83977      & 0.89984      & \textbf{0.91014}       \\
\multicolumn{1}{l|}{}                                   & \multicolumn{1}{l|}{$\frac{F1}{\Delta MSE}\uparrow$} & -            & 13.785       & 19.905       & 28.340       & \textbf{107.312}       & -            & 12.826       & 19.086       & 67.556       & \textbf{115.062}        \\ \hline
\multicolumn{1}{l|}{\multirow{4}{*}{\textbf{ETTm2}}}    & \multicolumn{1}{l|}{\textbf{MSE}$\downarrow$}    & 0.17298      & 0.22683      & 0.21571      & 0.21955      & \textbf{0.18256}       & 0.18318      & 0.23184      & 0.22649      & 0.19511      & \textbf{0.19199}       \\
\multicolumn{1}{l|}{}                                   & \multicolumn{1}{l|}{\textbf{MAE}$\downarrow$}    & 0.25810      & 0.32784      & 0.33267      & 0.32324      & \textbf{0.27607}       & 0.26881      & 0.33303      & 0.34109      & 0.28334      & \textbf{0.28276}       \\
\multicolumn{1}{l|}{}                                   & \multicolumn{1}{l|}{\textbf{F1}$\uparrow$}     & -            & 0.82208      & 0.65093      & 0.89852      & \textbf{0.89878}       & -            & 0.85668      & 0.65594      & 0.95990      & \textbf{0.98439}       \\
\multicolumn{1}{l|}{}                                   & \multicolumn{1}{l|}{$\frac{F1}{\Delta MSE}\uparrow$} & -            & 15.266       & 15.234       & 19.294       & \textbf{93.818}        & -            & 17.605       & 15.145       & 80.461       & \textbf{111.736}        \\ \hline
\multicolumn{1}{l|}{\multirow{4}{*}{\textbf{Electricity}}}      & \multicolumn{1}{l|}{\textbf{MSE}$\downarrow$}    & 0.14495      & 0.19987      & 0.18465      & 0.15820      & \textbf{0.15109}       & 0.17167      & 0.22530      & 0.21741      & 0.18121      & \textbf{0.18106}       \\
\multicolumn{1}{l|}{}                                   & \multicolumn{1}{l|}{\textbf{MAE}$\downarrow$}    & 0.24334      & 0.31041      & 0.30689      & 0.24149      & \textbf{0.25345}       & 0.26546      & 0.32839      & 0.33211      & 0.27967      & \textbf{0.27681}       \\
\multicolumn{1}{l|}{}                                   & \multicolumn{1}{l|}{\textbf{F1}$\uparrow$}     & -            & 0.71945      & 0.77701      & 0.75460      & \textbf{0.82032}       & -            & 0.72525      & 0.78445      & 0.76918      & \textbf{0.80495}       \\
\multicolumn{1}{l|}{}                                   & \multicolumn{1}{l|}{$\frac{F1}{\Delta MSE}\uparrow$} & -            & 13.100       & 19.572       & 56.951       & \textbf{133.603}       & -            & 13.523       & 17.150       & 80.626       & \textbf{85.724}        \\ \hline
\multicolumn{1}{l|}{\multirow{4}{*}{\textbf{Weather}}}  & \multicolumn{1}{l|}{\textbf{MSE}$\downarrow$}    & 0.15868      & 0.21131      & 0.20121      & 0.21888      & \textbf{0.16902}       & 0.17124      & 0.22702      & 0.21497      & 0.18223      & \textbf{0.17741}       \\
\multicolumn{1}{l|}{}                                   & \multicolumn{1}{l|}{\textbf{MAE}$\downarrow$}    & 0.20917      & 0.28469      & 0.29810      & 0.30013      & \textbf{0.23265}       & 0.21578      & 0.29392      & 0.30662      & 0.21994      & \textbf{0.22888}       \\
\multicolumn{1}{l|}{}                                   & \multicolumn{1}{l|}{\textbf{F1}$\uparrow$}     & -            & 0.88866      & 0.90706      & 0.92488      & \textbf{0.93083}       & -            & 0.91420      & 0.90971      & 0.95449      & \textbf{0.99475}       \\
\multicolumn{1}{l|}{}                                   & \multicolumn{1}{l|}{$\frac{F1}{\Delta MSE}\uparrow$} & -            & 16.885       & 21.328       & 15.363       & \textbf{90.022}        & -            & 16.389       & 20.803       & 86.851       & \textbf{161.224}        \\ \hline
\multicolumn{1}{l|}{\multirow{4}{*}{\textbf{Exchange}}} & \multicolumn{1}{l|}{\textbf{MSE}$\downarrow$}    & 0.10554      & 0.14682      & 0.14438      & 0.13774      & \textbf{0.11581}       & 0.09699      & 0.13924      & 0.13929      & 0.12004      & \textbf{0.10731}       \\
\multicolumn{1}{l|}{}                                   & \multicolumn{1}{l|}{\textbf{MAE}$\downarrow$}    & 0.22810      & 0.27998      & 0.29370      & 0.26911      & \textbf{0.24266}       & 0.22079      & 0.27481      & 0.28994      & 0.24886      & \textbf{0.23655}       \\
\multicolumn{1}{l|}{}                                   & \multicolumn{1}{l|}{\textbf{F1}$\uparrow$}     & -            & 0.91207      & 0.61842      & 0.96035      & \textbf{0.96741}       & -            & 0.94026      & 0.61816      & \textbf{0.95463}      & 0.95206       \\
\multicolumn{1}{l|}{}                                   & \multicolumn{1}{l|}{$\frac{F1}{\Delta MSE}\uparrow$} & -            & 22.095       & 15.922       & 29.825       & \textbf{94.198}        & -            & 22.255       & 14.614       & 41.416       & \textbf{92.254}       \\ \hline
\end{tabular}
\end{table*}

\section{Experiments}\label{sec_experiments}
In this section, we conduct experiments to answer the following research questions (RQs):
\begin{itemize}
  \item \textbf{RQ1.} How is the detectability and utility of the watermarked outputs?
  \item \textbf{RQ2.} How is the traceability of the watermarks against illegal knowledge distillation?
  \item \textbf{RQ3.} How does \modelname benefit from each key design?
  \item \textbf{RQ4.} How does the key hyper-parameters influence \modelname's performance?
  \item \textbf{RQ5.} A case study for watermarked outputs.
\end{itemize}

\subsection{Datasets}
We extensively assess our proposed \modelname on seven real-world datasets, covering energy, weather, and finance areas: ETTh1, ETTh2, ETTm1, ETTm2~\cite{zhou2021informer}, Electricity, Exchange~\cite{lai2018modeling}, and Weather.
All the datasets are publicly available in the popular unified time series library\footnote{\url{https://github.com/thuml/Time-Series-Library}}.
The detailed statistics introduction of these datasets is in Appendix~\ref{apd_dataset_details}.

\subsection{Evaluation Metrics}
Following the most common settings~\cite{zhou2021informer,cao2024tempo,liu2024unitime}, we use Mean Squared Error (MSE) and Mean Absolute Error (MAE) as the primary metrics to evaluate the utility of the time series output.
We use F1 scores to assess the detectability of watermarks.
Besides, we use the fraction of F1 scores and MSE $\frac{F1}{\Delta MSE}$, where $\Delta MSE = MSE_{watermarked} - MSE_{original}$ to straightforwardly show how ``effective'' the watermark algorithm in achieving high detectability within minor performance drop, where the higher $\frac{F1}{\Delta MSE}$ indicates better watermark algorithm.

\subsection{Baselines}
In this paper, we focus on two representative LLM-based forecasting models (TEMPO~\cite{cao2024tempo} and UniTime~\cite{liu2024unitime}) as backbones because they are widely recognized, state-of-the-art, and capture distinct methodological paradigms in the field. As most existing LLMTS models share a similar framework with these two, experiments on them already provide solid insights and yield a clearer evaluation. Although an exhaustive comparison with all existing methods might seem more comprehensive, it would largely introduce redundancy without offering additional meaningful conclusions.

We compare \modelname against several baselines:
(1) Clean, which represents the unwatermarked forecasting results;
(2) HTW \cite{Schaik2025Robust}, the only existing watermarking method specifically designed for \llmts;
(3) Forecasting Signal Watermarking (FSW), inspired by watermarking techniques from the image and multimedia domains \cite{badshah2016watermark,cox1997secure} and adapted to the unique characteristics of time series data. It embeds watermarks by directly modulating selected segments of the forecast signal in the time domain using predefined patterns (e.g., sine waves);
(4) \modelname(Soft), a variant of our method that replaces the hard-constrained PGD optimization with a simpler normalization-based “soft” constrained optimization (E.q.~\ref{eq_optim_with_norm}).

\subsection{Implementation Details}
We conduct our watermark experiments with two open-sourced and popular \llmts, TEMPO\footnote{\url{https://github.com/DC-research/TEMPO}} and UniTime\footnote{\url{https://github.com/liuxu77/UniTime}}.
Since TEMPO released a pre-trained checkpoint, we directly use this checkpoint for experiments.
While for UniTime, we train it on the seven datasets using its default setting to get a pre-trained \llmts model for watermark experiments.
For \modelname, by default, we set the number of watermarked patches $\alpha$ to $1$, the noise constraint to $0.01$, and the noise learning step to $0.1$, using the Adam optimizer. In Section~\ref{sec_hyper_param}, we further analyze the impact of these key hyperparameters. The Z-score threshold for watermark detection is set to $2$.

\subsection{Effectiveness of Watermarking (RQ1)}\label{sec_rq1}
In this section, we evaluate the effectiveness of the proposed watermarking algorithm. An effective watermark must strike a balance between detectability, which is the ability to reliably detect the watermark, and imperceptibility, which is to minimize any negative impact on the model's predictive performance. The comparison results between \modelname and several baselines are presented in Table~\ref{tb_watermark_validate}.

Our experimental setup is as follows. We begin by using pretrained \llmts models (UniTime and TEMPO) to generate forecasts on the test sets for each dataset. Next, we apply each watermarking algorithm to the model outputs, which are then mixed with the original (clean) outputs. The task for each watermarking method is to correctly identify which outputs have been watermarked within this mixed dataset.

The results shown in Table~\ref{tb_watermark_validate} indicate that \modelname introduces minimal degradation to the model's forecasting accuracy, especially when compared to the baselines. For example, on the ETTh1 dataset, HTW and FSW lead to increases in MSE of 14\% and 10.5\%, respectively, while \modelname results in only a 2\% increase. At the same time, \modelname consistently achieves the highest F1 scores across most settings, demonstrating superior watermark detection capability. Additionally, the variant \modelname(Soft), which replaces the hard-constrained PGD optimization with a simpler normalization-based ``soft'' constraint, yields comparable but slightly worse F1 scores. 
This is because, the normalization term will impede the noise's alignment optimization objective.
Meanwhile, the MSE costs for \modelname(Soft) are also slightly higher than \modelname, since normalization cannot guaranttee a hard constraint.
As a result, unlike prior methods that often sacrifice substantial output quality for detectability, \modelname ensures that perturbations remain within a bounded noise region, preserving output fidelity.

To more clearly illustrate the trade-off between detection effectiveness and model performance degradation, we report the ratio $\frac{\text{F1}}{\Delta \text{MSE}}$ in Table~\ref{tb_watermark_validate}. As the table shows, \modelname achieves up to five times higher scores on this metric compared to the baselines, further emphasizing its effectiveness in maintaining both robustness and imperceptibility.

Finally, the superiority of \modelname holds consistently across both TEMPO and UniTime and across all evaluated datasets, demonstrating the generalizability and versatility of our approach.

\begin{figure*}[!htbp]
  \centering
  \subfloat[F1 Scores with TEMPO.]{\includegraphics[width=0.49\textwidth]{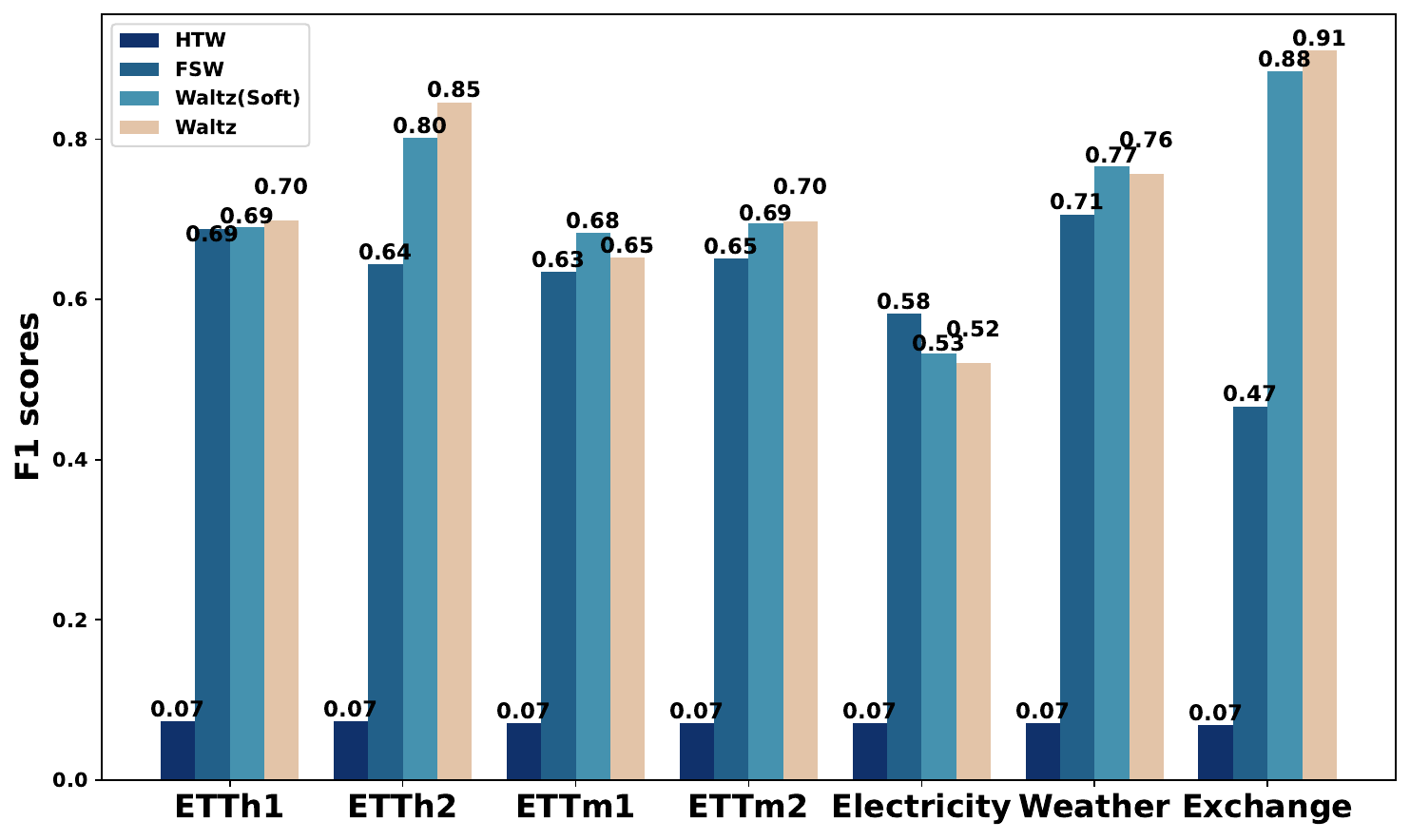}\label{fig_distill_tempo}}
  \hfil
  \subfloat[F1 Scores with UniTime]{\includegraphics[width=0.49\textwidth]{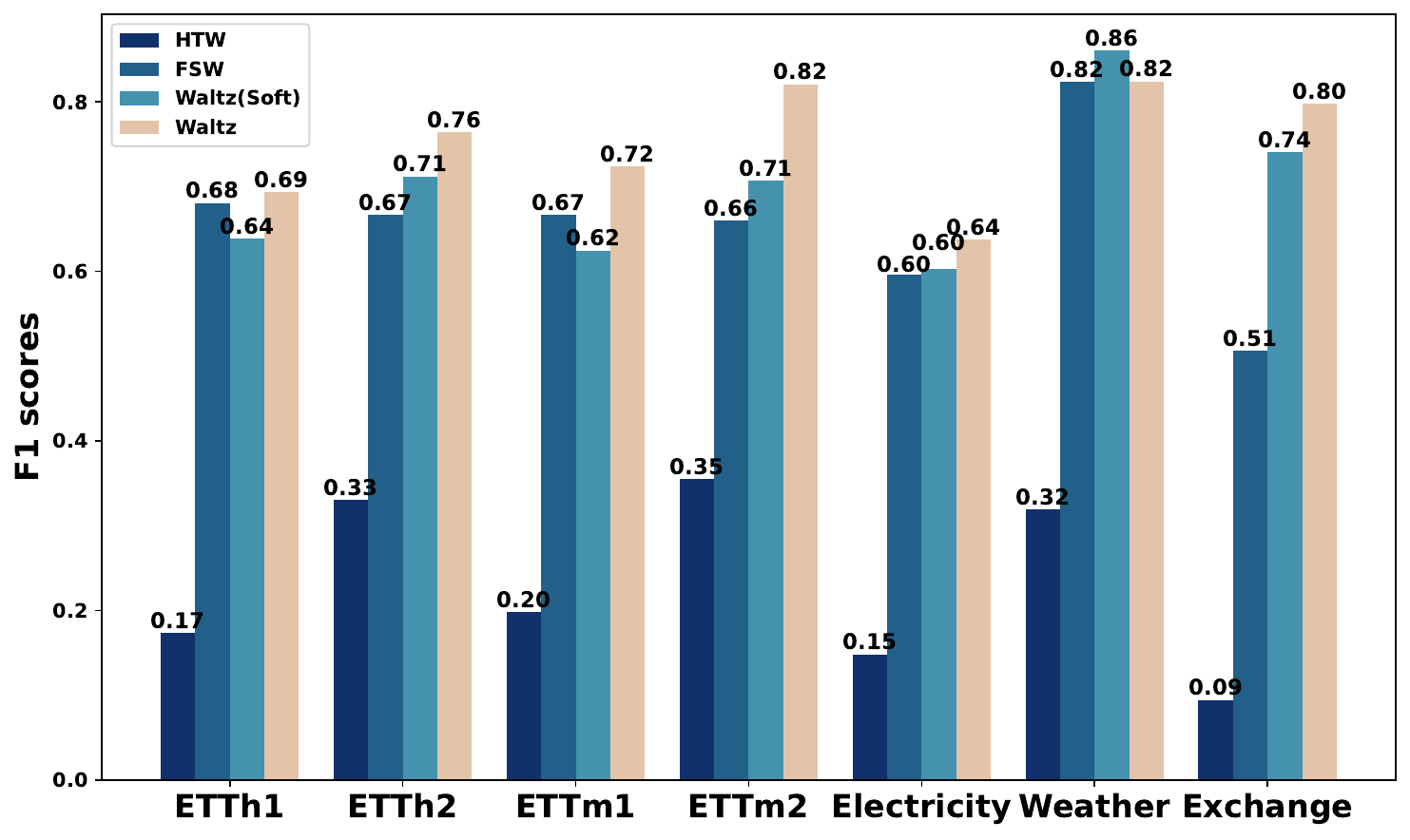}\label{fig_distill_unitime}}
  \caption{The traceability of watermark after knowledge distillation.}\label{fig_distill}
  \vspace{-8pt}
\end{figure*}

\subsection{Traceability of Watermarking (RQ2)}
An ideal watermark should not only be imperceptible and detectable but also traceable and robust in scenarios involving unauthorized use. In this section, we evaluate the traceability of \modelname under a practical threat model: unauthorized model distillation~\cite{peng2023you}. While some image watermarking studies also consider attacks such as offsetting, random cropping, and min-max insertion~\cite{yang2024gaussian}, we argue that such manipulations are impractical in the context of time series data. This is because these operations would severely distort the temporal dependencies and semantic patterns intrinsic to time series, rendering the data unusable.

To simulate a distillation attack, we first use the pretrained \llmts to generate forecasts on the training set of each dataset, applying watermarking during inference to produce a watermarked training dataset $\hat{\mathcal{D}}_{wm}$. 
We then assume an adversary uses $\hat{\mathcal{D}}_{wm}$ to train a new time series forecasting model from scratch. 
The goal is to assess whether the outputs from this distilled model still contain the watermark, i.e., whether a watermark detector can still identify them as watermarked.
For this evaluation, we adopt DLinear~\cite{zeng2023transformers} as the base model due to its favorable trade-off between training efficiency, forecasting accuracy, and widespread use in the literature.

Figure~\ref{fig_distill} presents the F1 scores of the distilled models across datasets. The results show that the watermark embedded by HTW is not preserved through the distillation process, with F1 scores dropping to near zero. This likely stems from the nature of HTW, which introduces visibly unnatural patterns into the forecasted outputs (as visualized in Figure~\ref{fig_case_study}) that deviate significantly from normal time series behavior, making them easy for a new model to disregard or ``unlearn'' during training.
FSW retains some watermark signal post-distillation, however, as discussed in Section~\ref{sec_rq1}, it substantially degrades output fidelity, limiting its practical usability.
In contrast, both \modelname(Soft) and \modelname consistently achieve strong F1 scores across most scenarios, demonstrating their robust traceability under distillation-based attacks. This suggests that our watermarking approach is not only minimally invasive but also resilient, maintaining watermark detectability even when the watermarked outputs are used to train a new model.

\subsection{Ablation Study (RQ3)} 
\modelname incorporates three key components in its watermarking and detection design: the use of cold tokens for watermarking, similarity-based insertion of watermark signals, and z-score-based detection. In this section, we conduct an ablation study to assess the necessity of each of these components. Specifically, we individually replace each design with a simpler, naïve alternative, and present the resulting performance in Table~\ref{tb_ablation_study}.
The variant labeled ``-cold token'' replaces cold tokens with randomly selected tokens. ``-similarity-based insertion'' refers to inserting watermark noise at randomly chosen positions rather than those selected based on patch similarity. Lastly, ``-z-score detection'' replaces our statistical detection method with a simpler top-k approach.
Since all variants are still optimized using PGD, ensuring that the magnitude of the watermark noise remains constrained, their impact differences on output quality is negligible. Therefore, we focus solely on F1 scores in Table~\ref{tb_ablation_study}.
Besides, we only display the results with TEMPO on four datasets due to the space limitation, similar conclusion can be obtained from other cases.

The results clearly show that removing any of these three carefully designed components leads to a substantial decline in F1 scores across all four datasets. 
For instance, replacing cold tokens with random tokens results in a sharp drop in F1 score on the Exchange dataset, from 0.96 to 0.71. 
Similarly, when watermark noise is inserted into randomly selected patches instead of using similarity-based positions, the F1 score on ETTh1 falls from 0.80 to 0.70. As for detection, using z-score detection proves significantly more effective than the top-k method, with F1 scores dropping by at least 0.15 when z-score detection is removed.

\begin{table}[]
  \caption{Ablation study. The F1 scores with TEMPO on four datasets. Same conclusion can be obtained from other datasets.} \label{tb_ablation_study}
  \resizebox{0.47\textwidth}{!}{
\begin{tabular}{l|cccc}
\hline
                              & \textbf{ETTh1}   & \textbf{ETTh2}   & \textbf{Weather} & \textbf{Exchange} \\ \hline
\textbf{\modelname} & \textbf{0.88600} & \textbf{0.91793} & \textbf{0.93083} & \textbf{0.96741}  \\
\textbf{-cold token}      & 0.72668          & 0.84162          & 0.80516          & 0.71664           \\ 
\textbf{-similarity-based insertion}            & 0.70716          & 0.83627          & 0.87015          & 0.93094           \\
\textbf{-z-score detection}           & 0.62560          & 0.70403          & 0.72052                & 0.80571           \\\hline
\end{tabular}}
\end{table}

\subsection{Hyper-parameter Analysis (RQ4)}\label{sec_hyper_param}
In this part, we analyze the impact of two key hyperparameters in \modelname: the noise constraint $\eta$ and the number of watermarked patches $\alpha$.

\begin{figure}[!htbp]
  \centering
  \subfloat[The performance trend with the number of watermarked patches $\alpha$.]{\includegraphics[width=0.48\textwidth]{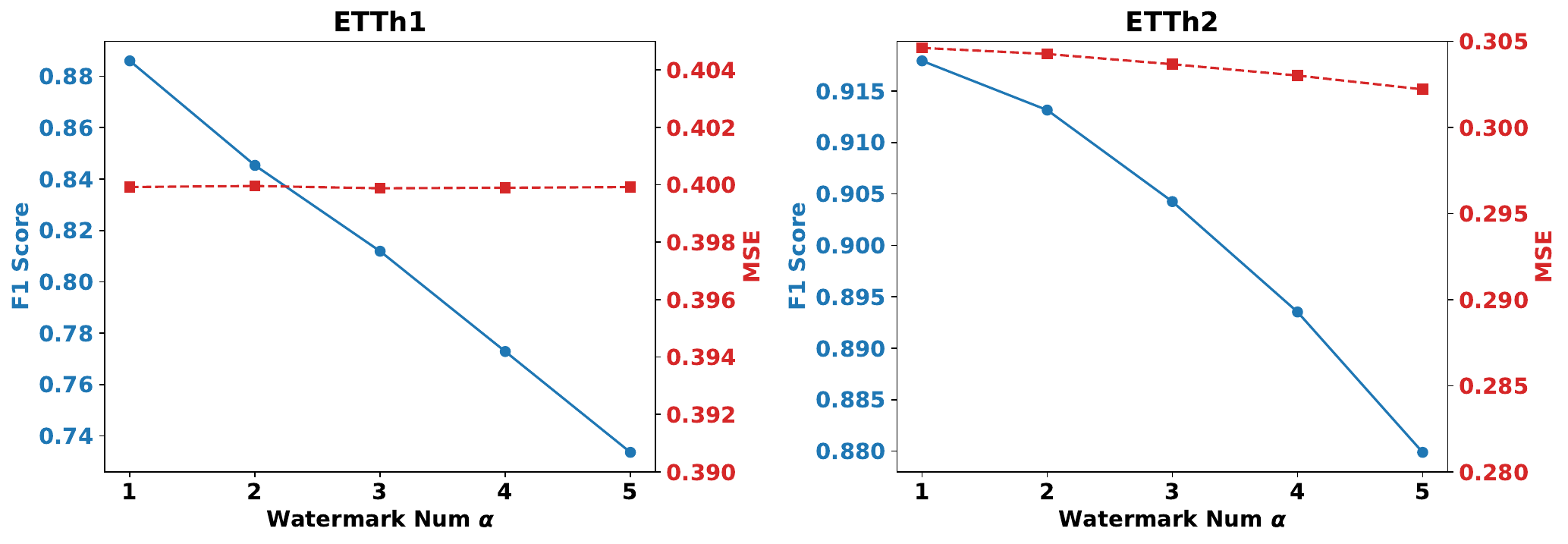}\label{fig_hyp_water}}
  \hfil
  \subfloat[The performance trend with noise scales $\eta$.]{\includegraphics[width=0.48\textwidth]{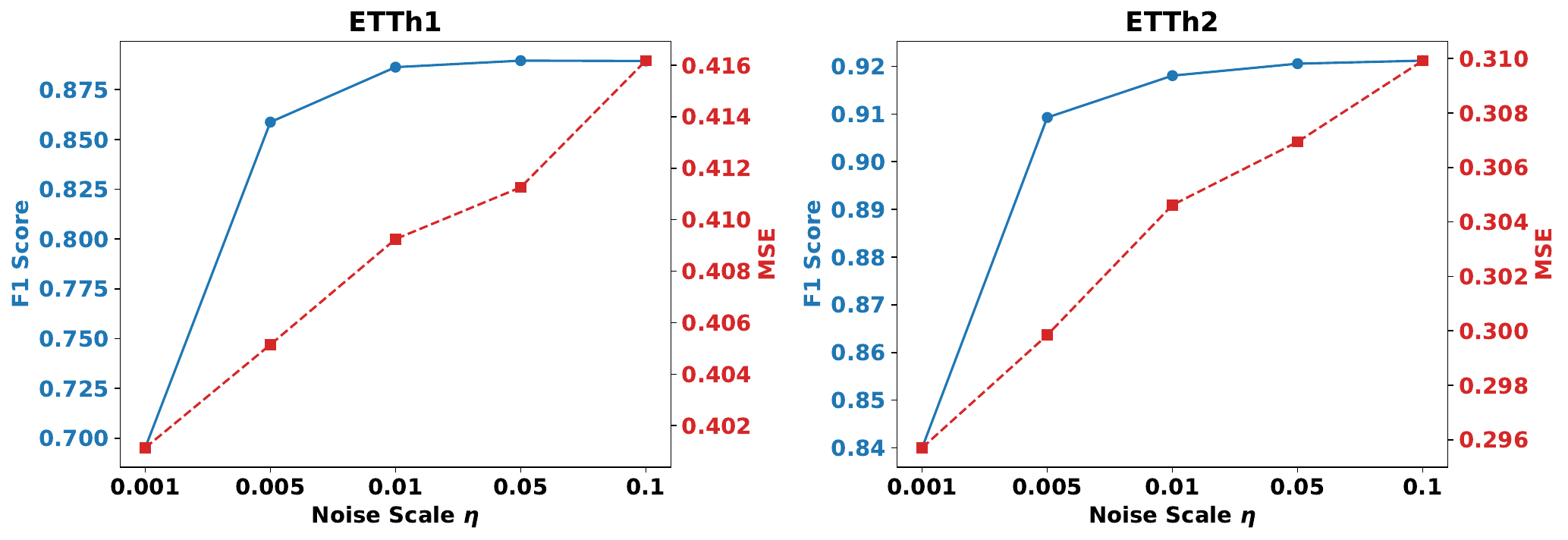}\label{fig_hyper_epsilon}}
  \caption{Hyper parameter analysis for the number of watermarked patches and the noise scales on ETTh1 and ETTh2 with TEMPO. Same observation can be found in other cases.}\label{fig_hyper} 
  \vspace{-8pt}
\end{figure}
\noindent\textbf{The Number of Watermarked Patch} $\alpha$.
As shown in Figure~\ref{fig_hyp_water}, increasing the number of watermarked patches from 1 to 5 leads to a noticeable decrease in F1 scores. This result is intuitive. Since the noise constraint $\eta$ remains fixed, as evidenced by the nearly unchanged MSE indicated by the red line in the figure, increasing the number of patches causes the watermarking noise to be spread more thinly across them. 
As a result, averagely, each individual patch receives a smaller perturbation, weakening the influence of the watermark on each patch embedding and preventing it from achieving the intended ``rewiring'' objective. 

\noindent\textbf{The Value of Noise Scale Constraint} $\eta$.  
Figure~\ref{fig_hyper_epsilon} illustrates how varying the noise constraint $\eta$ affects watermarking performance. 
We observe that loosening the constraint, i.e., increasing $\eta$, initially results in a rapid improvement in F1 scores, until $\eta = 0.01$. 
Beyond this point, further increases in noise scale yield only marginal gains in detection performance. 
Besides, the F1 score gains come at the cost of reduced output quality, as the watermarked predictions begin to diverge more noticeably from clean outputs. 
This highlights a clear and unavoidable trade-off between watermark detection accuracy and model utility: relaxing the noise constraint improves traceability but degrades the usability of the watermarked outputs.

\begin{figure}[!htbp]
  \centering
  \subfloat[TEMPO on ETTh1.]{\includegraphics[width=0.48\textwidth]{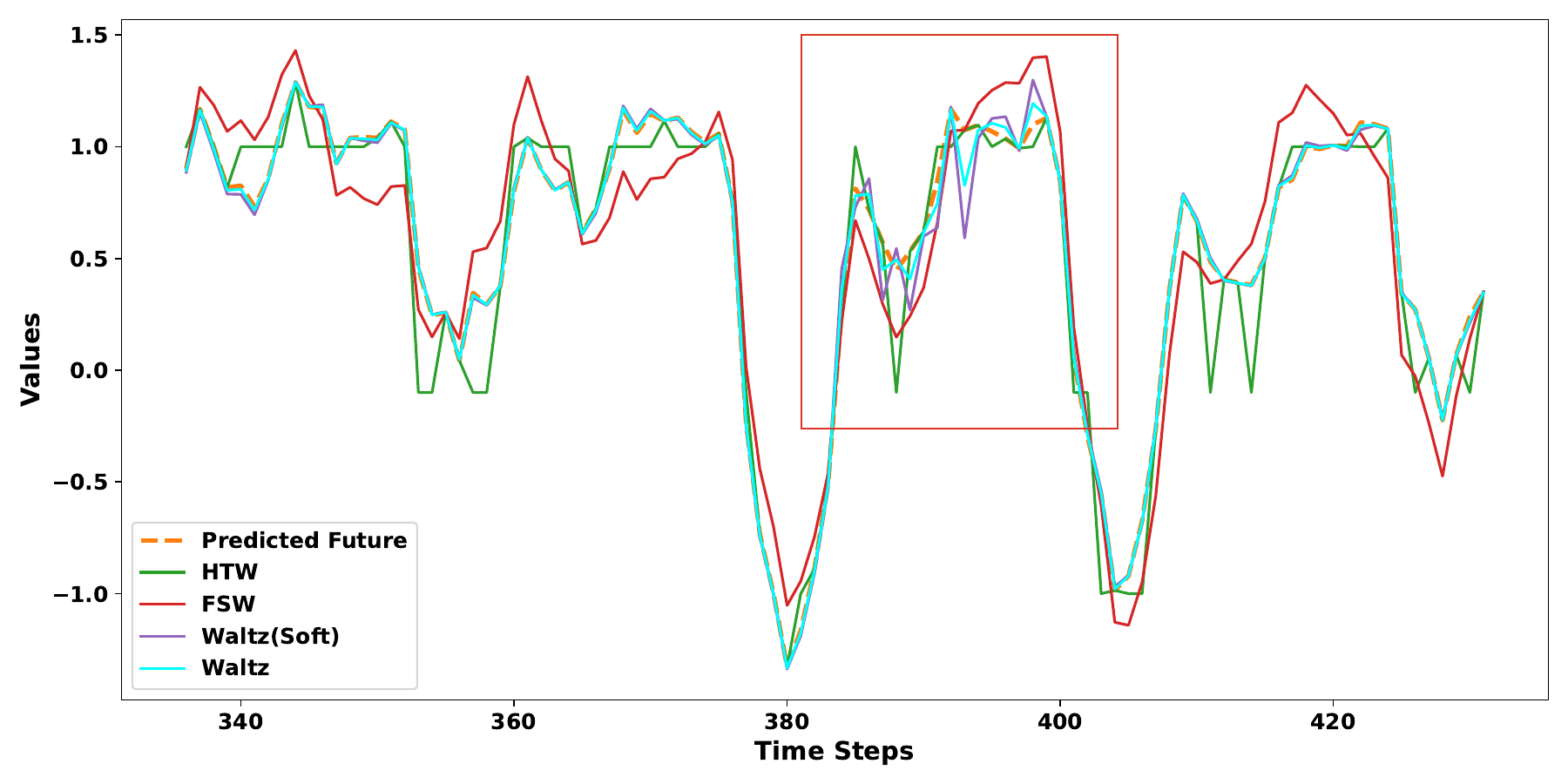}\label{fig_case_tempo_etth1}}
  \hfil
  \subfloat[TEMPO on ETTh2]{\includegraphics[width=0.48\textwidth]{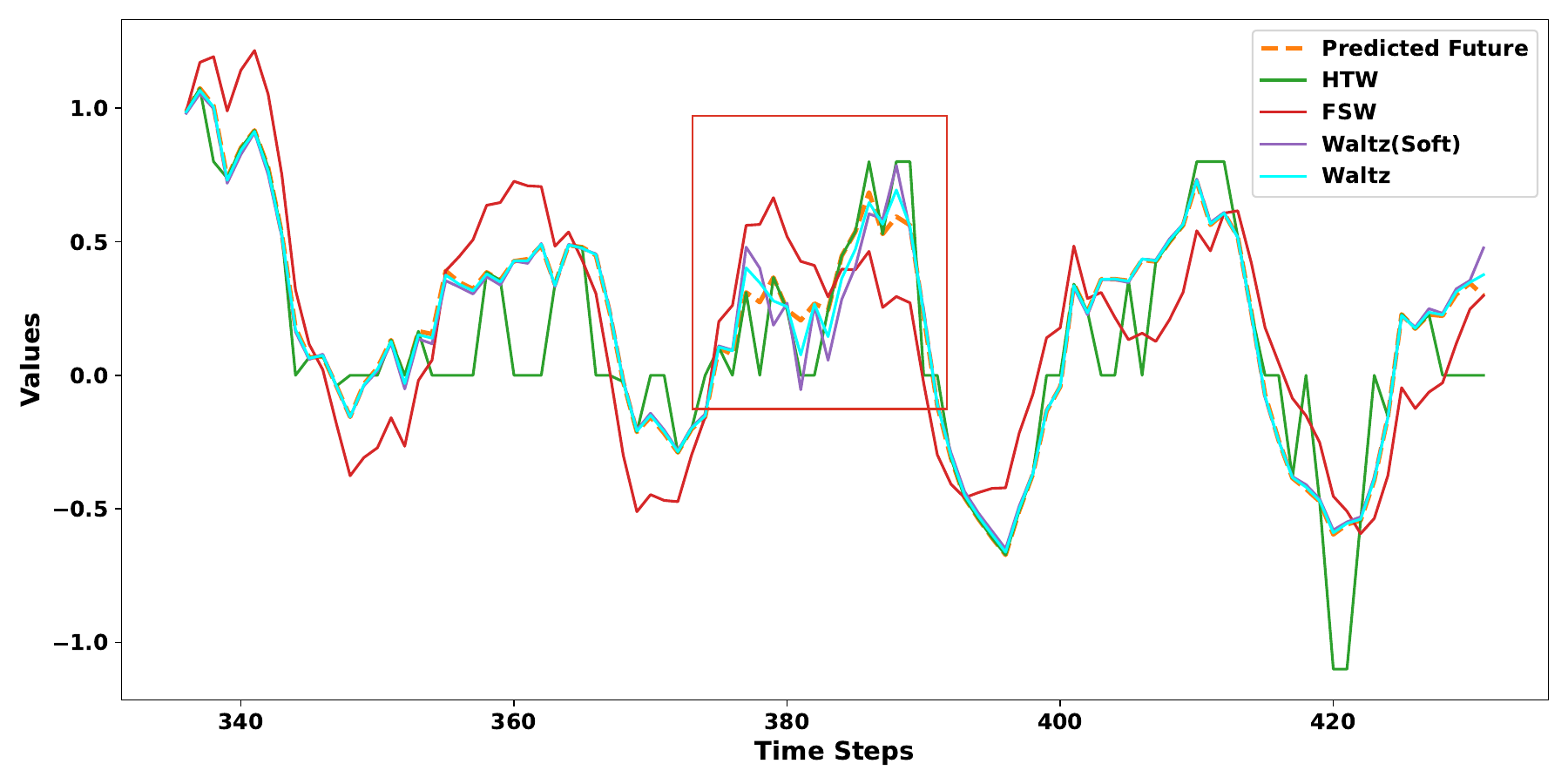}\label{fig_case_tempo_etth2}} 
  \caption{Case study for watermarked outputs.}\label{fig_case_study}  
  \vspace{-8pt}
\end{figure}
\subsection{Case Study (RQ5)}\label{sec_case_study}
Figure~\ref{fig_case_study} presents two illustrative examples from the ETTh1 and ETTh2 datasets, comparing the effects of different watermarking algorithms. The "Predicted Future" refers to the original, unwatermarked outputs generated by TEMPO. As observed, the baseline methods, HTW and FSW, introduce substantial alterations to the entire predicted time series, visibly distorting the forecasted patterns. In contrast, \modelname modifies only a small segment of the output (highlighted by the red box in the figure), and the changes are limited to minor value deviations within that localized patch. This demonstrates the strong imperceptibility of \modelname, as it embeds the watermark without compromising the overall structure or quality of the prediction.

\section{Conclusion}\label{sec_conclusion}
In this paper, we propose an effective watermarking framework, \modelname, designed to protect the intellectual property of \llmts and prevent misuse of their generated outputs. \modelname embeds watermarks by rewiring the alignment between time series embeddings and the LLM's cold token embeddings. To ensure imperceptibility, a projected gradient descent (PGD) method is employed to enforce a hard constraint on the magnitude of the watermarking noise. Additionally, a z-score-based detection mechanism is introduced to validate the presence of the watermark.
Extensive experiments conducted on seven datasets using two representative \llmts demonstrate that \modelname achieves high detection accuracy and remains traceable even after unauthorized distillation. Importantly, these results are obtained without significantly compromising the output quality of the protected models.

\bibliographystyle{ACM-Reference-Format}
\bibliography{sample-base}

\appendix


\begin{table*}[!h]
\caption{The statistics of datasets.}\label{tb_dataset_statistics}
\begin{tabular}{l|ccccc}
\hline
\textbf{Dataset}     & \textbf{Variable} & \textbf{Frequency} & \textbf{Length} & \textbf{Domain} & \textbf{Start Time} \\ \hline
\textbf{ETTh1/ETTh2} & 7                 & 1 hour             & 17,420          & Energy          & 2016/7/1            \\
\textbf{ETTm1/ETTm2} & 7                 & 15 min             & 69,680          & Energy          & 2016/7/1            \\
\textbf{Electricity} & 321               & 1 hour             & 26,304          & Energy          & 2012/1/1            \\
\textbf{Weather}     & 22                & 10 min             & 52,696          & Weather         & 2020/1/1            \\
\textbf{Exchange}    & 8                 & 1 day              & 7,588           & Finance         & 1990/1/1            \\ \hline
\end{tabular}
\end{table*}

\section{More Related Work}\label{apd_relatedwork}
In this section, we briefly review the literature on two closely related topics: LLM for time series forecasting and digital watermarking.
Other involved topics like time series forecasting can be referred to corresponding surveys~\cite{wang2024deep,jin2024survey,wen2023transformers}.

\subsection{LLM for Time Series Forecasting}
Due to the impressive generalization and reasoning capabilities of large language models (LLMs)~\cite{min2023recent}, there has been a growing interest in extending their applications to domains with distinct modalities and data structures, such as graph learning~\cite{ren2024survey}, tabular data~\cite{fang2024large}, and recommender systems~\cite{zhao2024recommender}. 
Inspired by these developments, researchers in time series forecasting have also begun to explore the potential of LLMs for time series modeling~\cite{zhang2024large, jin2024position}.

Broadly, two main research directions have emerged in the intersection of LLMs and time series forecasting. The first leverages LLMs to encode and understand textual information that can assist forecasting in multimodal time series tasks~\cite{zhong2025time, liu2024time, wang2024news, jia2024gpt4mts, kong2025time}. The second aims to directly utilize LLMs' robust sequential pattern recognition capabilities to improve forecasting performance on pure time series data. In this paper, we primarily focus on the latter, as it has shown promising empirical results, while the former remains in its early stages due to the scarcity of high-quality multimodal time series datasets.

A very early effort in this direction is by Gruver et al.~\cite{gruver2023large}, who demonstrated the zero-shot capability of LLMs on time series data.
However, their method naively represents time series as strings of numerical digits, resulting in limited performance compared to conventional forecasting models.
Zhou et al.~\cite{zhou2023one} provided theoretical evidence that the self-attention mechanism in LLMs can function similarly to principal component analysis, supporting the potential of LLMs in this domain. They also empirically showed that by applying a linear patching layer to project time series into token sequences, and fine-tuning only the positional embeddings and normalization layers, pretrained LLMs can achieve competitive forecasting performance compared to traditional deep learning models.
TEMPO~\cite{cao2024tempo} further advanced this line of work by decomposing time series into trend, seasonal, and residual components, encoding them via a linear encoder, and integrating them with prompts to feed into an LLM for sequence encoding. The output is then passed through a decoder to generate future forecasts. This framework has inspired many subsequent works, which typically follow a similar pipeline while introducing more sophisticated time series encoding and alignment techniques to boost performance.
For example, UniTime~\cite{liu2024unitime} uses attention-based encoding for time series patches and prompt-based mechanisms to enable cross-domain forecasting. 
Jin et al.~\cite{jintime} proposed a reprogramming method that aligns time series inputs with LLM word embeddings and enhances instruction prompts to improve adaptation.
Pan et al.~\cite{pan2024s} introduced semantically-informed prompt learning, aligning the pretrained LLM semantic space with time series embeddings to enable forecasting in a shared representation space.
In summary, the general framework adopted by most existing \llmts approaches is illustrated in the left part of Figure~\ref{fig_overview}. In this work, we design a post-hoc watermarking method tailored for this widely adopted \llmts framework, aiming to improve model IP protection without compromising forecasting performance.

\subsection{Digital Watermarking}
With the rapid advancement of large models, growing concerns have emerged around their potential misuse and the lack of intellectual property (IP) protection, both of which could pose significant risks to society.
Digital watermarking~\cite{cox2002digital}, which embeds subtle signals into multimedia content to make it traceable, has become a widely adopted technique to mitigate these risks.
In computer vision, for example, Wen et al.~\cite{wen2023tree} proposed Tree-ring, which embeds watermarks into images by injecting noise in the Fourier space to prevent the unauthorized use of large diffusion models. 
In natural language processing, Kirchenbauer et al.~\cite{kirchenbauer2023watermark} introduced a decoding-time watermarking method that subtly biases token sampling by increasing the likelihood of randomly selected “green” tokens, producing statistically detectable patterns without degrading output quality.
And many variants have been proposed based on its basic idea to watermark LLMs~\cite{liang2024watermarking,zhang2024remark}.
Although \llmts models leverage LLMs for time series forecasting, existing watermarking methods designed for original text-based LLMs are not directly applicable due to key differences in data structure and model architecture. 
Specifically, (1) Text consists of discrete tokens, while time series data is continuous and numerical, which means time series data has infinite ``vocabulary'', making token-level watermarking approaches inapplicable.
(2) \llmts models incorporate specialized time series encoders and decoders, which fundamentally alter the input-output pipeline of the base LLM. As a result, typical watermarking methods that inject signals during training or at the token decoding stage cannot be readily adopted.

In the time series forecasting domain, watermarking remains relatively underexplored.
TimeWak~\cite{soi2025timewak} introduces a watermarking strategy specifically for diffusion-based time series generation, but it is incompatible with the \llmts paradigm.
To date, HTW~\cite{Schaik2025Robust} is the only watermarking method designed for \llmts, but its approach is rudimentary and imposes a significant degradation in forecasting accuracy, as shown in Table~\ref{tb_watermark_validate}.
In this work, we propose a more effective and principled watermarking approach tailored for \llmts. Our method embeds robust and detectable watermarks while preserving forecasting utility, addressing the limitations of prior techniques.

\section{Details of Datasets}\label{apd_dataset_details}
In this paper, we select seven datasets that lie at the intersection of those used in the original papers of the two baselines, TEMPO and UniTime, for presentation convenience. 
Table~\ref{tb_dataset_statistics} summarizes the statistics of these datasets.
ETTh1 and ETTh2 record seven parameters from electrical transformers on an hourly basis over a two-year period, yielding a total of 17,420 time steps starting from July 1, 2016. ETTm1 and ETTm2 similarly record transformer parameters but at 15-minute intervals. Electricity contains hourly electricity consumption data from 321 clients, spanning from 2012 to 2014. Weather includes meteorological factors collected every 10 minutes from 22 stations, starting on January 1, 2020. Exchange consists of daily exchange rates of eight countries' currencies against the US dollar, covering the period from 1990 to 2010, with a total of 7,588 time steps.
Note that, for TEMPO, since the final model checkpoint is publicly available, we use it directly as the base model for our watermarking experiments to reduce the cost of training LLM-based time series models, which is not the primary focus of this paper. 
In contrast, for UniTime, we train the model on these datasets ourselves, as no pre-trained checkpoint is provided. The training, validation, and test splits follow the original paper~\cite{liu2024unitime}.









\end{document}